\documentclass[traditabstract,twocolumn]{aa}

\usepackage{epsfig,graphicx,natbib}
\usepackage{longtable}

\def\PKS1830{\hbox{PKS\,1830$-$211}}
\def\kms{\hbox{km\,s$^{-1}$}}
\def\cm-2{\hbox{cm$^{-2}$}}
\def\Tcmb{\hbox{$T_\mathrm{CMB}$}}
\def\aaa{\hbox{H$_2^{35}$Cl$^+$}}
\def\bbb{\hbox{H$_2^{37}$Cl$^+$}}
\def\ccc{\hbox{H$_2$Cl$^+$}}

\begin{document}

\title{Detection of chloronium and measurement of the $^{35}$Cl/$^{37}$Cl isotopic ratio at $z$=0.89 toward PKS\,1830$-$211}

\author{S. Muller \inst{1}
\and J.\,H.~Black \inst{1}
\and M.~Gu\'elin \inst{2,3}
\and C.~Henkel \inst{4,5}
\and F.~Combes \inst{6}
\and M.~G\'erin \inst{3}
\and S.~Aalto \inst{1}
\and A.~Beelen \inst{7}
\and J.~Darling \inst{8}
\and C.~Horellou \inst{1}
\and S.~Mart\'in \inst{2}
\and K.\,M.~Menten \inst{4}
\and Dinh-V-Trung \inst{9}
\and M.\,A.~Zwaan \inst{10}
}
\institute{
Department of Earth and Space Sciences, Chalmers University of Technology, Onsala Space Observatory, SE-43992 Onsala, Sweden
\and Institut de Radioastronomie Millim\'etrique, 300, rue de la piscine, 38406 St Martin d'H\`eres, France 
\and LRA/LERMA, CNRS UMR 8112, Observatoire de Paris \& Ecole Normale Sup\'erieure, Paris, France
\and Max-Planck-Institut f\"ur Radioastonomie, Auf dem H\"ugel 69, D-53121 Bonn, Germany
\and Astron. Dept., King Abdulaziz University, P.O. Box 80203, Jeddah, Saudi Arabia
\and Observatoire de Paris, LERMA, CNRS, 61 Av. de l'Observatoire, 75014 Paris, France
\and Institut d'Astrophysique Spatiale, B\^at. 121, Universit\'e Paris-Sud, 91405 Orsay Cedex, France
\and Center for Astrophysics and Space Astronomy, Department of Astrophysical and Planetary Sciences, University of Colorado, 389 UCB, Boulder, CO 80309-0389, USA
\and Institute of Physics, Vietnam Academy of Science and Technology, 10 DaoTan, ThuLe, BaDinh, Hanoi, Vietnam
\and European Southern Observatory, Karl-Schwarzschild-Str. 2, 85748 Garching b. M\"unchen, Germany
}

\date {Received  / Accepted}

\titlerunning{H$_2$Cl$^+$ toward \PKS1830}
\authorrunning{Muller et al. 2014}

\abstract{We report the first extragalactic detection of chloronium (H$_2$Cl$^+$), in the $z$=0.89 absorber in front of the lensed blazar \PKS1830. The ion is detected through its 1$_{11}$-0$_{00}$ line along two independent lines of sight toward the North-East and South-West images of the blazar. The relative abundance of \ccc\ is significantly higher (by a factor $\sim$7) in the NE line of sight, which has a lower H$_2$/H fraction, indicating that \ccc\ preferably traces the diffuse gas component. From the ratio of the \aaa\ and \bbb\ absorptions toward the SW image, we measure a $^{35}$Cl/$^{37}$Cl isotopic ratio of $3.1_{-0.2}^{+0.3}$ at $z$=0.89, similar to that observed in the Galaxy and the solar system.
}
\keywords{quasars: absorption lines - quasars: individual: \PKS1830\ - galaxies: ISM - galaxies: abundances - ISM: molecules - radio lines: galaxies
}
\maketitle

\section{Introduction}

A plethora of new interstellar molecules, notably simple hydrides, has been discovered as a result of the recent opening of the submillimeter window, from space with the Herschel Space Observatory or from the ground, for example with the Atacama Pathfinder EXperiment (APEX) and Caltech Submillimeter Observatory (CSO) telescopes. Hydrides (that is, molecular species composed of a single heavy element with one or more hydrogen atoms) are formed by the first chemical reactions in the atomic gas component, and are therefore at the basis of interstellar chemistry. They are powerful probes of the interstellar environment and offer a variety of astrophysical diagnostics (e.g., \citealt{qin10,ger10,god12,fla13,sch14}).

One such hydride is chloronium, \ccc, which was first detected by \cite{lis10} in foreground absorption toward the sources NGC\,6334I and Sgr\,B2(S) with the Herschel Space Observatory. \cite{neu12} extended observations of \ccc\ to six Galactic sources, four in absorption and two in emission (toward OMC\,1: Orion Bar and Orion South). These constitute the only observations of chloronium in the literature to date. The other chlorine-bearing molecules detected in the interstellar medium are hydrogen chloride, HCl (\citealt{bla85}) and the chloroniumyl ion, HCl$^+$ (\citealt{luc12}), while metal halides such as NaCl, AlCl, and KCl were detected in the circumstellar envelope IRC+10216 (\citealt{cer87}) and in the atmosphere of Io (\citealt{lel03,mou10,mou13}).

Here, we report the first extragalactic detection of chloronium, in the $z$=0.89 absorber toward the $z$=2.5 blazar \PKS1830, and a measurement of the $^{35}$Cl/$^{37}$Cl isotopic ratio at a look-back time of more than half the present age of the Universe.

\section{Data}

The 1$_{11}$-0$_{00}$ line of the para spin-species of both the \aaa\ and \bbb\ isotopologues, with rest frequencies of $\sim$485.4\,GHz and $\sim$484.2\,GHz, respectively, was detected in absorption at $z_{abs}$=0.89 (i.e., redshifted to $\sim$257\,GHz) toward the blazar \PKS1830\ with the Atacama Large Millimeter/submillimeter Array (ALMA). The observations and details of the data reduction are described by \cite{mul14} (hereafter {\em Paper\,I}). We used ALMA Band~6 data from four observing runs performed between April and June 2012. The total resulting on-source integration time was approximately 30\,min. The two lensed images of the background blazar, separated by 1$''$, were resolved by the ALMA array, but each remained a point-like source. One absorption spectrum was extracted toward each image with the task {\sc uvmultifit} (\citealt{mar14}), by modelling the visibilities with two point-like sources where the relative positions were fixed and the flux densities left as free parameters. The two chloronium isotopologues were observed simultaneously in the same spectral window, 1.875\,GHz wide and with a spectral channel spacing of 0.488\,MHz. The resulting velocity resolution of the chloronium spectra is 1.2\,\kms\ after Hanning smoothing.

The chloronium frequencies and relative intensities of the hyperfine components are taken from the work by \cite{ara01}. The dipole moment, $\mu$=1.89\,D, is from an ab initio calculation by \cite{holger08}. All velocities are referred to redshift $z$=0.88582 in the heliocentric frame. 

\begin{figure}[ht!] \begin{center}
\includegraphics[width=8.8cm]{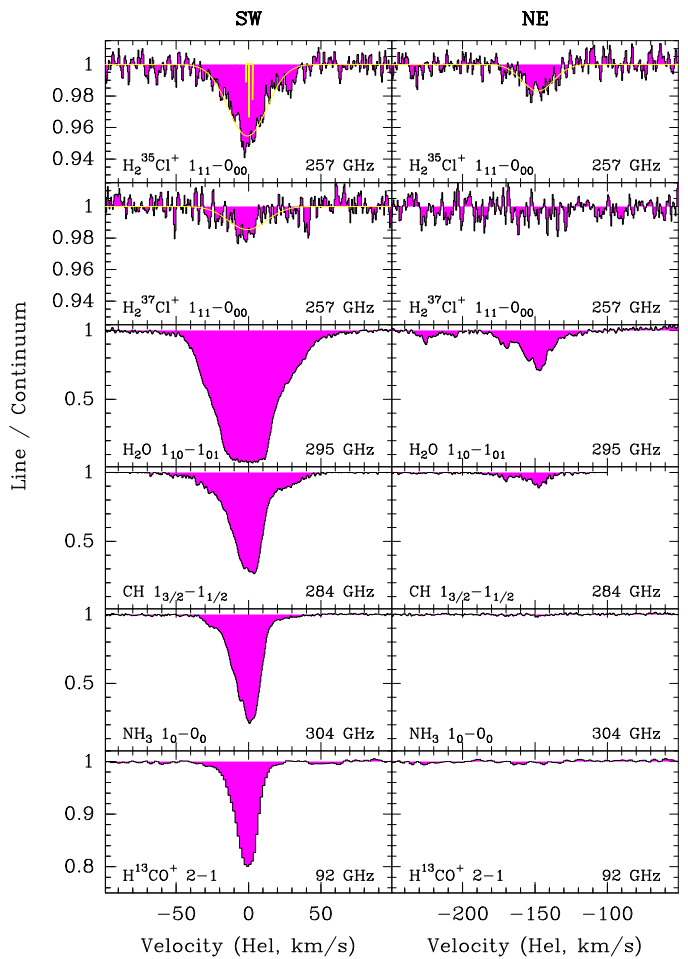}
\caption{Spectra of the \aaa\ and  \bbb\  1$_{11}$-0$_{00}$ (para) line and other species, all observed by ALMA between April and June 2012, toward the \PKS1830\,SW image (left) and the NE image (right). The hyperfine structure for the \aaa\ para-line is shown (top-left). That for \bbb\ is similar. The Gaussian fits are overlaid in light yellow. The redshifted line frequency is given in the bottom-right corner of each box. A detailed presentation of the ALMA data is given by \cite{mul14}.}
\label{fig:spec}
\end{center} \end{figure}

The absorption of \aaa\ is detected toward both images of the blazar while that of \bbb\ is detected only toward the SW image (see Fig.\,\ref{fig:spec}). The lines are shallow and optically thin toward both images, absorbing only a few percent of the continuum background. Because the width of the absorption profile is larger than the splitting of the hyperfine structure (over $\sim$4.5\,\kms, as shown in Fig.\,\ref{fig:spec}), we did not deconvolve the spectra.

\section{Discussion}

\subsection{Column densities and abundances}

Chloronium is a widespread species in the Galactic diffuse medium (\citealt{neu12}), and it is not surprising to detect it in the SW absorption toward \PKS1830. Indeed, the SW line of sight is particularly rich in molecules, with more than 40 species detected to date (\citealt{mul11} and {\em Paper\,I}). What is surprising at first glance, however, is to detect chloronium absorption toward the NE image, with a SW/NE absorption depth ratio of only 3--4, while all other molecular species observed so far have a much deeper absorption toward the SW image, with SW/NE abundance ratios of a few tens (e.g., \citealt{mul11}). This is well illustrated in Fig.\,\ref{fig:spec} by the comparison of the chloronium absorption toward both images with the absorption from species such as H$_2$O, CH, NH$_3$, and H$^{13}$CO$^+$. All spectra were observed by ALMA between April and June 2012 and are not affected by time variations (see \citealt{mul08} and the discussion in {\em Paper\,I}). Note that the H$_2$O absorption is heavily saturated toward the SW image. Only the H\,I line (e.g., \citealt{koo05}) shows an absorption deeper (by a factor $\sim$2) toward the NE image than toward the SW image.

The line profile of the \aaa\ absorption toward the SW image is wider (FWHM=32$\pm$1\,\kms) than that of the optically thin H$^{13}$CO$^+$ 2-1 line (FWHM=17.1$\pm$0.3\,\kms). In particular, the \aaa\ absorption shows an additional weak feature at a velocity of $\sim$30\,\kms, where the H$_2$O and CH profiles have a prominent line wing, which most likely represents a diffuse gas component (see the discussion in {\em Paper\,I}).

We estimate an integrated opacity of $\sim$1.5\,\kms\ along the SW line of sight for the \aaa-para line. Assuming a rotation temperature locked to the temperature of the cosmic microwave background, \Tcmb=5.14\,K at $z$=0.89 (see \citealt{mul13}), a source-covering factor $f_c$ of unity, and an ortho/para ratio of 3 (\citealt{ger13}), we derive a column density of $\sim$1.4$\times$10$^{13}$\,\cm-2. In fact, the covering factor of the SW image is not unity, but $\sim$95\%, as shown by the saturation level of the 557\,GHz water line (see {\em Paper\,I}). However, this does not introduce a noticeable difference in the apparent opacity of \ccc\ since the line is optically thin and $f_c$$\sim$1. With the same assumptions along the NE line of sight, we estimate an integrated opacity of $\sim$0.4\,\kms, corresponding to a \aaa\ column density of 4$\times$10$^{12}$\,\cm-2. With total H$_2$ column densities of 2$\times$10$^{22}$ and 1$\times$10$^{21}$\,\cm-2\ along the SW and NE lines of sight, respectively (\citealt{mul11} and {\em Paper\,I}), we finally derive fractional abundances of [\aaa]/[H$_2$]$\sim$6$\times$10$^{-10}$ (SW) and $\sim$4$\times$10$^{-9}$ (NE), that is, a \ccc\ abundance relative to H$_2$ $\sim$7 higher along the NE line of sight. Note that the covering factor is not well known toward the NE image, but the ALMA data suggest 0.3$<$$f_c$$<$1.0. Assuming $f_c$$<1$ would increase the true opacity, column density, and relative abundance of \ccc, and give an even higher relative abundance ratio than for the SW line of sight.

The chemistry of interstellar chlorine is thought to be simple and well understood (see \citealt{neu09}); but the observed abundances of the ions HCl$^+$ and \ccc\ in the Galactic interstellar medium are rather higher than predicted in current models (\citealt{neu12}). In diffuse clouds, the chemistry starts from ionized chlorine (the first ionization potential of chlorine, 12.97\,eV, is slightly lower than that of hydrogen), forming HCl$^+$ by reaction with H$_2$. A further reaction of HCl$^+$ with H$_2$ leads to \ccc. The molecule can in turn react with free electrons (dissociative recombination) to form HCl or Cl. In dense clouds, the chemistry is driven by cosmic-ray ionization and not by UV-photoionization, and neutral chlorine can react with the H$_3^+$ ion to form HCl$^+$, which again can react with H$_2$ to produce \ccc. The chemical rates and balance of the above reactions are not precisely known, but the relative abundance of \ccc\ clearly depends on the ionization level of chlorine, that is, on the UV irradiation field and atomic hydrogen density (\citealt{neu09}). 

The significantly higher relative abundance of \ccc\ in the NE line of sight, where the absorbing gas has a lower molecular fraction (H$_2$/H) than in the SW, confirms that the chloronium abundance is enhanced in the diffuse, more atomic, interstellar component (\citealt{neu12}).

\subsection{$^{35}$Cl/$^{37}$Cl isotopic ratio at $z$=0.89}

The clear detection of both \aaa\ and \bbb\ isotopologues toward the SW image allows us to measure their abundance ratio. The simple, well-known chlorine chemistry and the most likely weak fractionation between both isotopologues ensure that this ratio reflects the $^{35}$Cl/$^{37}$Cl isotopic ratio. Note that both isotopologues were observed simultaneously and within the same 1.875\,GHz-wide spectral window, which minimizes instrumental uncertainties. From a simultaneous fit of the SW spectrum with a single Gaussian component (centroid and width constrained to the same values for both isotopologues), we derive an isotopic ratio $^{35}$Cl/$^{37}$Cl of $3.1_{-0.2}^{+0.3}$, where uncertainties correspond to a 68\% confidence level from a Monte Carlo analysis. If the ratio is the same toward the NE image, \bbb\ should be just at the limit of detection. We estimate a lower limit of $^{35}$Cl/$^{37}$Cl$>$1.9 at a 99.7\% confidence level. Slightly deeper observations should thus allow us to measure the $^{35}$Cl/$^{37}$Cl ratio toward this component, which intercepts the absorber at a larger galactocentric radius ($\sim$4\,kpc {\em vs} $\sim$2\,kpc for the SW image). 


\begin{table}[ht]
\caption{Astronomical measurements of the $^{35}$Cl/$^{37}$Cl ratio.} \label{tab:isotopicratio}
\begin{center} \begin{tabular}{lccr}
\hline
\multicolumn{1}{c}{Source} & $^{35}$Cl/$^{37}$Cl & Species & Ref. \\
\hline
Solar system & 3.13 & Cl & 1  \\
\hline
IRC+10216 & $2.3\pm0.5$     & NaCl, AlCl & 2 \\
Ori\,A    & $\sim$4 -- 6        &  HCl & 3  \\
IRC+10216 & $3.1\pm0.6$ & NaCl,\,KCl,\,AlCl & 4 \\
IRC+10216 & $2.30\pm0.24$ & NaCl, AlCl & 5 \\
W3\,A $^\dagger$ & $2.1\pm 0.5$ & HCl & 6 \\
NGC\,6334I, Sgr\,B2(S)\,$^\dagger$ & $\sim$2.7 -- 3.3 & \ccc and HCl & 7 \\
10 Galactic sources  & $\sim$1 -- 5 $^\ddagger$ & HCl & 8 \\   
W31\,C, Sgr\,A $^\dagger$ & $\sim$2 -- 4 & \ccc &9 \\
W31\,C $^\dagger$ & $2.1 \pm 1.5$ & HCl$^+$ & 10 \\
W31\,C $^\dagger$ & $\sim$2.9 & HCl & 11 \\
CRL\,2136 & $2.3\pm0.4$ $^\diamond$ & HCl & 12 \\
\hline
\PKS1830(SW) $^\dagger$ & $3.1_{-0.2}^{+0.3}$ & \ccc & 13 \\
\PKS1830(NE) $^\dagger$ &  $>$1.9 $^*$ & \ccc & 13 \\
\hline 
\end{tabular} \tablefoot{$\dagger$ The line was detected in absorption against the background source.
$\ddagger$ Possible confusion between absorption and emission features. 
$\diamond$ A CH feature might partially blend with the H$^{37}$Cl signal, possibly affecting the derived value of the ratio.
$*$ At 99.7\% confidence level.\\
{\bf References:} (1) \cite{lod03}; 
(2) \cite{cer87}; (3) \cite{sal96}; (4) \cite{cer00}; (5) \cite{kah00}; (6) \cite{cer10}; (7) \cite{lis10}; (8) \cite{pen10}; (9) \cite{neu12}; (10) \cite{luc12}; (11) \cite{mon13}; (12) \cite{got13}; (13) this work.} 
\end{center} \end{table}

The measurement of $^{35}$Cl/$^{37}$Cl at $z$=0.89 (SW) is within the range of values found in Galactic sources (see Table\,\ref{tab:isotopicratio}), and is, in particular, identical to the terrestrial ratio within the uncertainty. In contrast to $^{35}$Cl/$^{37}$Cl, the isotopic ratios of $^{18}$O/$^{17}$O, $^{28}$Si/$^{29}$Si, and $^{32}$S/$^{34}$S in the same $z$=0.89 absorber (SW line of sight) were found to deviate significantly by factors of 2 -- 3 from their local Galactic values (see \citealt{mul06,mul11,mul13}). While little is known about the conditions of the $z$=0.89 absorber (metallicity, elemental abundances, star formation activity), its look-back time is more than half the present age of the Universe. Consequently, we expect that the interstellar enrichment is more dominated by nucleosynthesic products from massive stars, especially concerning heavy elements such as silicon, sulfur, and chlorine, than a region with a similar galactocentric radius in the Milky Way.

The two stable isotopes of chlorine can be produced during hydrostatic oxygen burning from the $\alpha$-elements $^{32}$S and $^{36}$Ar via $^{32}$S($\alpha$,$p$)$^{35}$Cl and $^{36}$Ar($n$,$\gamma$)$^{37}$Ar($\beta^+$)$^{37}$Cl, respectively (e.g., \citealt{thi85}). $^{37}$Cl can also be produced from $^{35}$Cl by $s$-process. $^{36}$Cl is unstable, but its half-lifetime is of about 3$\times$10$^5$\,yr, long enough to catch a second neutron to reach $^{37}$Cl before decay. In the interstellar gas, spallation reactions from cosmic rays on argon can also lead to chlorine isotopes.

\begin{figure*}[ht!] \begin{center}
\includegraphics[height=\textwidth,angle=270]{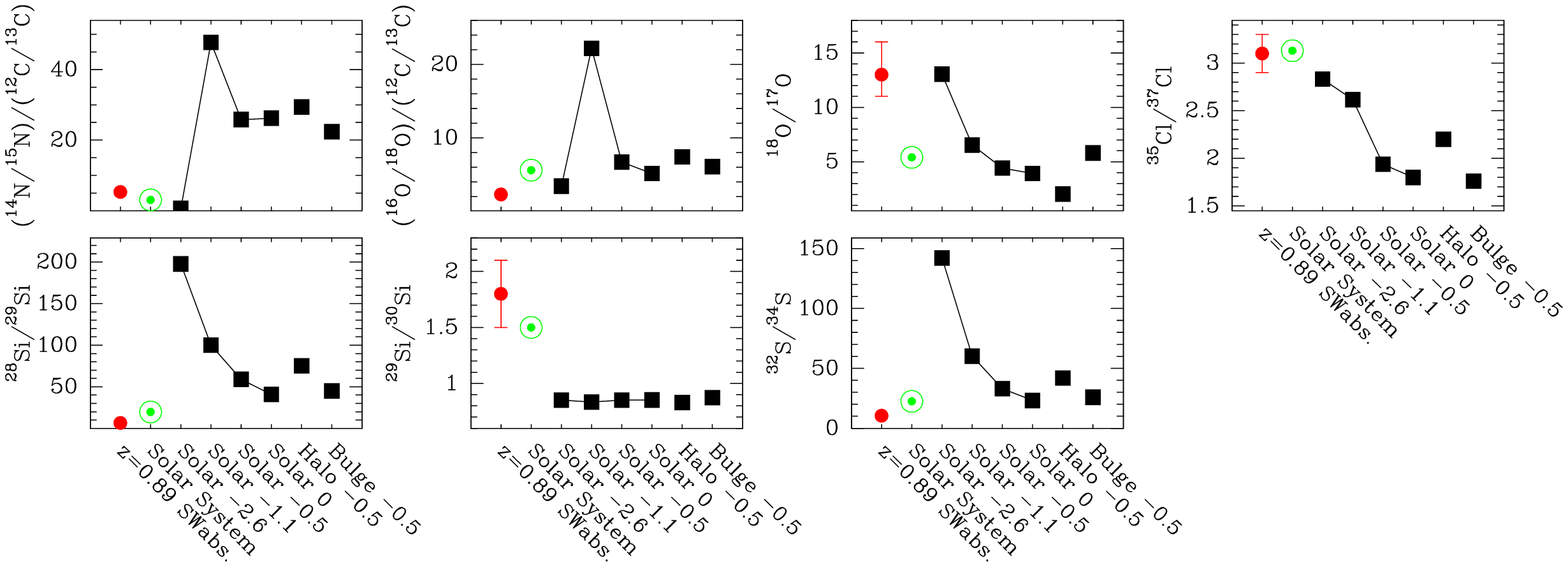}
\caption{Comparison of the isotopic ratios of C, N, O, S, Si, and Cl measured at $z$=0.89 toward \PKS1830(SW) (\citealt{mul06,mul11,mul13} and this work) and in the solar system ({\em solar symbols in green}, \citealt{lod03}), and predictions from evolution models from \cite{kob11} ({\em black squares}) for the solar neighbourhood (solar, at [Fe/H]= $-$2.6, $-$1.1, and $-$0.5), and halo and bulge at [Fe/H]=$-$0.5. The $^{14}$N/$^{15}$N and $^{16}$O/$^{18}$O ratios are normalized by the $^{12}$C/$^{13}$C ratio, because of the difficulties of measuring all three separately in the $z$=0.89 absorber toward \PKS1830.}
\label{fig:simul}
\end{center} \end{figure*}

In Fig.\,\ref{fig:simul}, we compare the isotopic ratios measured at $z$=0.89 toward \PKS1830(SW) with theoretical predictions of time/metallicity evolution models by \cite{kob11} in the Milky Way. For the solar neighbourhood, three epochs/metallicities are considered by \cite{kob11}: at [Fe/H]=$-$2.6 (metal-poor type-II supernovae, SNe\,II), [Fe/H]=$-$1.1 (SNe\,II + AGB stars), and [Fe/H]=$-$0.5 (SNe\,II + AGB + SNe\,Ia). Predictions at [Fe/H]=$-$0.5 for the halo and bulge components are also reported in the figure. The interstellar $^{12}$C/$^{13}$C, $^{14}$N/$^{15}$N, and $^{16}$O/$^{18}$O ratios are difficult to measure in general, mainly because of their relatively high values that result in either high opacity for lines of the most abundant isotopologues or sensitivity problems for lines of the rarest isotopologues. To alleviate these problems, we normalized the $^{14}$N/$^{15}$N and $^{16}$O/$^{18}$O ratios by $^{12}$C/$^{13}$C, considering the double-ratio obtained from, for example H$^{13}$CN/HC$^{15}$N or H$^{13}$CO$^+$/HC$^{18}$O$^+$ (see \citealt{mul11}). Hence, all the ratios for the $z$=0.89 absorber in Fig.\,\ref{fig:simul} are measured through optically thin lines and are therefore reliable. 

All the ratios measured at $z$=0.89 (SW), including $^{35}$Cl/$^{37}$Cl, agree very well with the predictions by \cite{kob11} for the solar neighbourhood at [Fe/H]=$-$2.6, except those of silicon and sulfur. This discrepancy should be viewed as an interesting constraint for chemical evolution models.


\section{Summary and conclusions} \label{Conclusion}

The chloronium ion, \ccc, was detected in the $z$=0.89 absorber toward the lensed blazar \PKS1830. The \ccc\ relative abundance along the NE line of sight was found to be enhanced by a factor $\sim$7 with respect to the SW line of sight. Since the NE line of sight is thought to be more diffuse, with a lower molecular gas fraction (H$_2$/H), this suggests that \ccc\ is a good tracer of the diffuse gas component. Toward the SW image, at a look-back time of more than half the present age of the Universe, we measured a $^{35}$Cl/$^{37}$Cl isotopic ratio of $3.1_{-0.2}^{+0.3}$, identical to its value in the solar system within the uncertainty, and within the range of values found in Galactic sources. Slightly deeper observations are expected to allow us to measure the $^{35}$Cl/$^{37}$Cl ratio in the NE line of sight, that is, at a larger galactocentric radius in the absorber, which will provide an additional interesting constraint for chemical evolution models.

The detection of \ccc\ toward \PKS1830\ suggests that other chlorine-bearing species might be easily detectable (e.g., with ALMA), in particular hydrogen chloride, HCl. Future observations of other hydrides, such as CH$^+$, OH$^+$, H$_2$O$^+$, HF, or ArH$^+$, will provide more information on the conditions in this (so far unique) extragalactic molecular absorber.

\begin{acknowledgement}
This paper makes use of the following ALMA data: ADS/JAO.ALMA\#2011.0.00405.S. ALMA is a partnership of ESO (representing its member states), NSF (USA) and NINS (Japan), together with NRC (Canada) and NSC and ASIAA (Taiwan), in cooperation with the Republic of Chile. The Joint ALMA Observatory is operated by ESO, AUI/NRAO and NAOJ. The financial support to Dinh-V-Trung from Vietnam's National Foundation for Science and Technology (NAFOSTED) under contract 103.08-2010.26 is gratefully acknowledged.
\end{acknowledgement}

\end{document}